# Principal Component Analysis for Nonlinear Optical Microscopic Chemical Imaging of Nitrogen Gas

**Logan Carlson[1], Devesh Bhattarai[1], Mamata Subedi[1], Haifeng Wang[2] and Gombojav O. Ariunbold[1, 3, 4, *]**

[1] Department of Physics and Astronomy, Mississippi State University, Starkville, MS 39762, USA; lac575@msstae.edu, db3079@msstae.edu, ms4532@msstate.edu, ag2372@msstae.edu

[2] Department of Industrial and Systems Engineering, Mississippi State University, Starkville, MS 39762, USA; wang@ise.msstate.edu

[3] Department of Physics, National University of Mongolia, Ulaanbaatar 210646, Mongolia

[4] Institute of Physics and Technology, Mongolian Academy of Sciences, Ulaanbaatar 13330, Mongolia

[*] Correspondence: ag2372@msstgate.edu

**Abstract:**

We have implemented principal component analysis for microscopic wide-field chemical imaging via coherent Raman spectroscopy. Microscopic imaging of nitrogen gas has been challenging due to extremely weak signals stemming from low order Raman interaction. Wide-field coherent Raman micro-spectroscopy has demonstrated the ability to chemically distinguish nitrogen gas although it has been difficult to quantify spatial-density information due to significant levels of background noise. By subtracting the Gaussian beam shape and removing contributions from uninformative noise simultaneously from the set of images, we can reconstruct the normalized intensity fluctuations. Our analysis demonstrates that nitrogen gas within microvolume can be rapidly monitored under ambient conditions in less than 0.2 seconds. We believe that our work has the potential to improve visualization of microscopic flows due to molecular dynamics of gases and/or liquids otherwise invisible to infrared optical techniques.

## 1. Introduction

Nitrogen has the highest concentration among gases in the atmosphere constituting 78% of the total volume. It plays a crucial role in the formation of basic ingredients of life such as enzymes and amino acids [1]. Industrial applications of nitrogen include welding, food packaging, and production of ammonia for fertilizers, etc. Various techniques such as electrochemical sensors, optical sensors and spontaneous Raman spectroscopy are adopted to detect nitrogen species in ambient air. However, there are unique limitations imposed on each detection technique due to several factors like the material composition, high operating temperatures, the concentration of gas being investigated, background fluorescence, etc. While non-dispersive infrared (NDIR) sensors can sense nitrogen compounds, they are ineffective in sensing nitrogen gas molecules considering they are nonpolar and inactive in the infrared region [2-5]. For purposes in research areas like chemical sensing, materials science and biological studies, optical microscopy techniques are effective tools in probing the state of gases.

Optical microscopic imaging techniques are further categorized into linear and nonlinear optical microscopic imaging based on light-matter interactions to create an image. In linear microscopy techniques, a change in the intensity of light causes a proportional change in the sample's response [6]. As for gases, a linear microscopy technique like confocal laser scanning microscopy can be applied for air density sensing by intentionally varying the ambient air pressure [7]; however, these are ineffective in single shot imaging of more complicated gas dynamics. Nonlinear microscopy techniques require high light intensity for interactions to occur, and the sample response is not proportional to light intensity [8]. Many non-linear microscopic imaging techniques remain unexplored for ambient gases, especially those that are inactive in the infrared region [9]. One of the prominent nonlinear optical microscopic imaging techniques is wide-field coherent anti-Stokes Raman spectroscopy (CARS) [10, 11] that uses three laser pulses to generate a highly sensitive, label-free chemical image of a sample by tuning two of the laser pulses to have a frequency difference resonant to a vibrational mode of the molecules being analyzed.

The coherent nature of CARS signals enables rapid imaging of the sample, and its high sensitivity allows for the detection of relatively weak signals [12, 13]. With gaseous samples, using the CARS techniques, it has been able to detect pyridine vapor at room temperature even at low concentrations as low as 30 ppm [11]. Due to its versatility, we have chosen wide-field CARS for chemical microimaging of nitrogen gas. As presented in our previous experiment [14], wide-field CARS allows us to selectively view the peak from Raman active vibrational mode for nitrogen gas at 2330 cm$^{-1}$. CARS technique overcomes the limitations of a traditional Raman spectroscopic technique by avoiding fluorescence and enhancing signal power via a coherent nonlinear four-wave mixing (FWM) process [15, 16]. FWM is a third-order nonlinear optical process that occurs when the refractive index of a medium depends on the intensity of light passing through it [17]. The efficiency of this process is governed by the phase matching relationship between the interacting waves [18]. In our previous work, wide-field CARS imaging enabled us to chemically distinguish nitrogen gas from background [15], but spatial-density information is still elusive. These challenges have motivated us to adopt an appropriate and popular machine learning model for background subtraction known as the principal component analysis method.

Machine learning models have been extensively implemented in the domain of physical science, medicine, engineering, etc. Notable applications of machine learning models include breast cancer prediction [19] and medical image recognition [20]. One of the widely implemented models for feature identification in machine learning is principal component analysis (PCA). PCA is a dimensionality reduction technique used to transform a dataset into a new orthogonal basis defined by the variance structure of the data itself [21]. The main purpose of this technique is to simplify complex datasets. Through this process, a possibly huge set of original and correlated variables are transformed into a smaller set of new and uncorrelated variables [22] known as the principal components. The order of these components depends on the amount of variance they capture, keeping the first few components that lead to maximum variance and discarding the rest [23]. As PCA compresses data by reducing its dimensionality [24], data analysis is particularly less overwhelming. It also helps reveal unusual patterns in data and facilitates visualization by projecting high-dimensional data onto the principal components as a scoring method [25].

PCA is increasingly useful for projecting spectra or image sequences into components that capture the most significant variations while reducing redundancy and noise [26]. Projecting the original data onto these eigenvectors yields component scores that represent how strongly each frame or spectrum expresses a given feature. This provides a compact and ordered description of the dataset in terms of the most significant statistical patterns. An important advantage of PCA is that the dataset can be reconstructed using only a subset of components. This enables targeted analysis such as background subtraction [27, 28], noise suppression [29], and visualization of specific spectral features [30]. Hence, PCA serves as a powerful tool for identifying correlated signal features, isolating physical dynamics, and distinguishing coherent emission from incoherent background contributions in the context of coherent spectroscopic imaging.

In this paper, we implement PCA to assist in removing background contributions from CARS signal of nitrogen gas molecules. This process is conducted by identifying and removing background features including consistent detector artefacts and intensity profile. The work is divided into sections as follows: the methodology of image recording is described in the next section. Then, the data analysis of the experimental results is performed in the following section. The concluding remarks and possible extensions to this work in future are discussed in the closing section.

## 2. Methodology

In our previous study, we have seen that single-frame CARS imaging allows for fast data collection, but the collected images still contain the Gaussian beam intensity profile [15]. The presence of background makes further analysis more difficult using traditional image analysis techniques. Thus, we have employed PCA to enhance data quality and extract meaningful signal components. PCA helps separate the real vibrational signal from background and random noise by reducing the dimension of the CARS data into independent components.

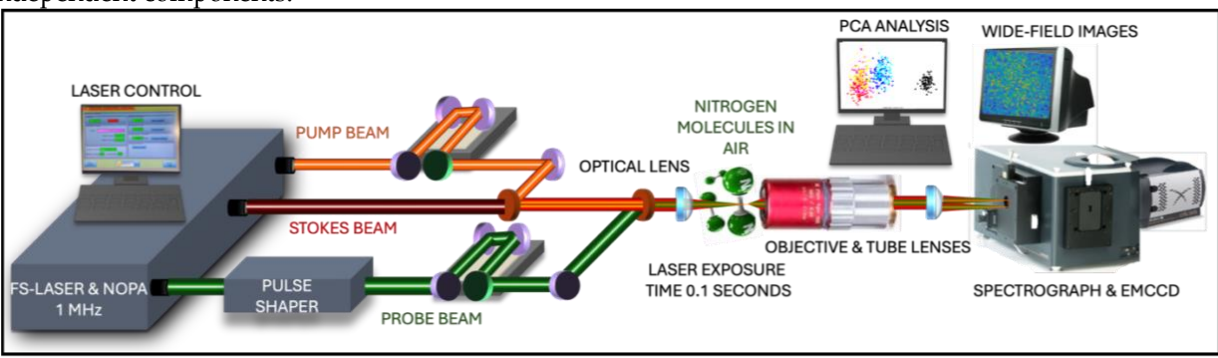

Figure 1. A schematic of the experimental imaging setup using the wide-field three color ultrafast CARS microscopy. Pump, Stokes, and Probe beams are produced by the femtosecond laser (MXR – Clark) combined with the NOPA. Data is recorded using a spectrograph with imaging EMCCD. Images are captured by a lab computer equipped with software to control the EMCCD camera and spectrograph shutter. Principal component analysis is performed for image data. Laser exposure times are 0.1 and 0.2 seconds. A spectral filter is used in front of open slit of the spectrograph (not shown).

For this study, the data is collected using the same methodology in [15], with the same pump, probe, and Stokes beams and atmospheric conditions of the sample volume. A schematic of the experimental setup is shown in Figure 1. When all three beams (pump, Stokes, and probe beams) are present, a wide-field CARS signal is observed that forms an image at the center with a full width at half maxima (FWHM) of 45 μm. Raw CARS spectral data is captured using an electron-multiplying charge-coupled device (EMCCD) camera. Wide-field CARS images as well as the background are collected at integration times of 0.1 and 0.2 s. A total of ninety-nine frames are recorded for each scan set out of total 6 scans. All three beams for CARS signal are present in scans 1, 2 and 3 with an exposure time of 0.2 s per frame. For scans 4 and 5, all three beams for CARS signal are present but have an exposure time of 0.1 s per frame. However, scan 6 only contains the probe beam with an exposure time of 0.1 s per frame with all other beams blocked. The kinetic image sequence provides a visualization of CARS signal from nitrogen gas as shown in our previous work [15].

The obtained images are saved and then cropped down either to 30 μm by 30 μm or 100 μm by 100 μm sets about the peak of the beam with a pixel size of 0.75μm x 0.75μm. The cropped image data consisting of multiple frames recorded in kinetic mode for six scans is then structured into a comprehensive data matrix, with each image reshaped into a vector represented by a row of the matrix. In this case, each row represents an image from one of the scans, and the columns represent specific pixel locations across all images. The resulting data matrix then undergoes a normalization process before PCA by applying the MATLAB z-score command. This process subtracts the mean intensity of each image and divides by its standard deviation. The ultimate effect of this normalization is to provide the CARS signal scans as well as background scans in terms of their variance, which normalizes them to a similar scale irrespective of the integration times.

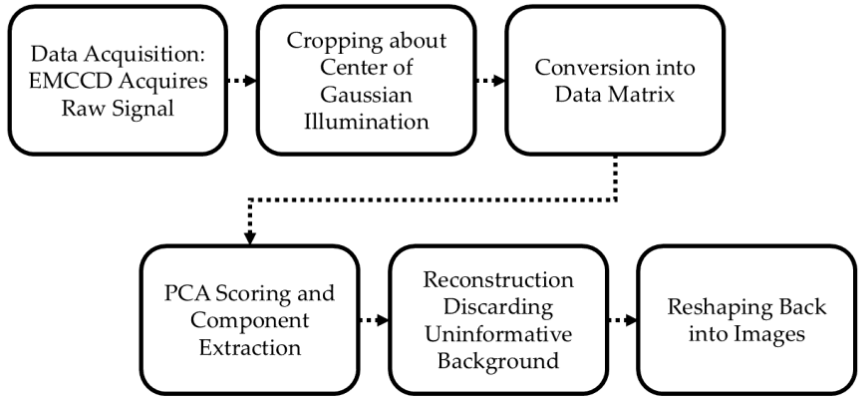

Figure 2: A flowchart diagram describing the entire process from data collection to obtaining signal dominant components. The process begins with data collection, which is then cropped about an approximate center of an illumination profile. The cropped images are then reshaped into a two-dimensional matrix and decomposed into a score matrix and principal component matrix via PCA. The uninformative background is then inspected and discarded during reconstruction of the data matrix. This reconstructed matrix is then reshaped back into the final reconstructed set of images.

The data is then processed using PCA to separate the primary CARS signal from background variations, thereby improving image contrast and accuracy. This analysis uses the MATLAB PCA command to decompose the dataset into coefficients and scores matrices. The coefficients capture spatial features (eigenvectors) while the scores track likeness to these features across all frames. Finally, the original matrix is then reconstructed from a subset of the decomposed principal components to understand the key patterns extracted on each sample. A flowchart in Figure 2 summarizes the entire process from data acquisition to obtaining final reconstructed CARS images. It is also necessary to note that the resulting scans contain sources of variance such as dark current and read noise within the EMCCD detector in addition to shot noise from the signal.

## 3. Results and Analysis

From the original 6 recorded scan sets, the images were cropped to two sizes – 30 μm by 30 μm and 100 μm by 100 μm with a pixel size of $0.75 \mu m^2$. We proceed by investigating the effectiveness of the background subtraction technique for each of the crop sizes when the illumination profile's FWHM is detected to be 45μm in our previous work. [15] For each of the six scan sets, each set of cropped frames is normalized to limit the influence of integration time in the images during PCA. Normalization is performed by subtracting the mean intensity from each image and dividing by the standard deviation of each scan set using the MATLAB z-score command. This process ensures the deviations are similar across both scans of 0.2 s and 0.1 s.

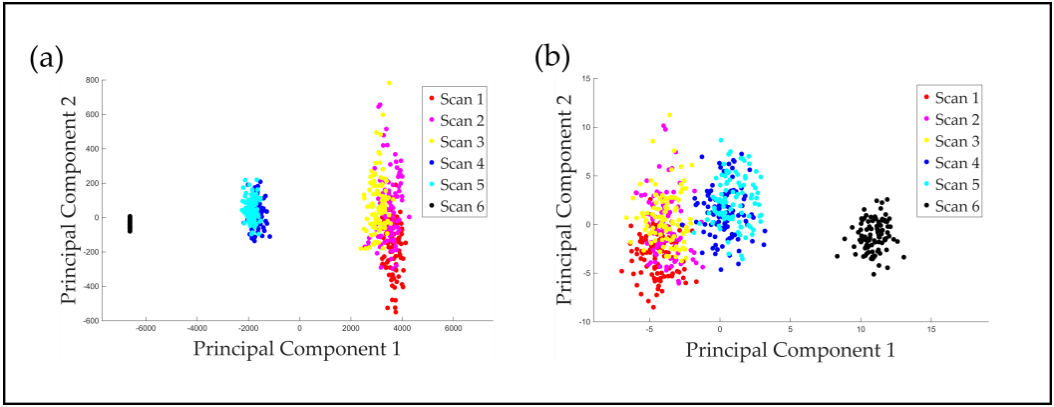

Figure 3: Plots representing the principal component scores of each scan: (a) before normalization and (b) after normalization. They are colored by scan set, with the Scan 6 being colored in black, since it is not CARS signal. In the unnormalized images, scan sets are distinguished by integration time which corresponds to the illumination profile shown in principal component 1. The normalized images cause the CARS scans to become much more similar, which removes extreme cases such as the 0.2s integration time scans and Scan 6 which only contains the probe beam.

The normalization process is followed by PCA procedure, in which the features are extracted from the data matrix and scores are computed for each image and feature in the set. We note that scan set 6 where only the probe beam is present is included in the data matrix, so the background is more distinguished and subtracted than without it, and PCA can be aware of the variance structure of the background. Next, we investigate the PC scores of each scan group before and after normalization to observe its effect. Figure 3 illustrates the principal component score plot of each scan group before and after normalization. In Figure 3 (a), each scan group is distinguished by the principal component 1 score, which correlates to the integration time and peak intensity of the scan set. However, in Figure 3 (b), the nitrogen images in which CARS signal is present (scans 1, 2, 3, 4, and 5) are more similar but they are still distinguished from the background without CARS signal in scan 6, indicating that normalization helps in standardizing the signal and distinguishing from background contributions.

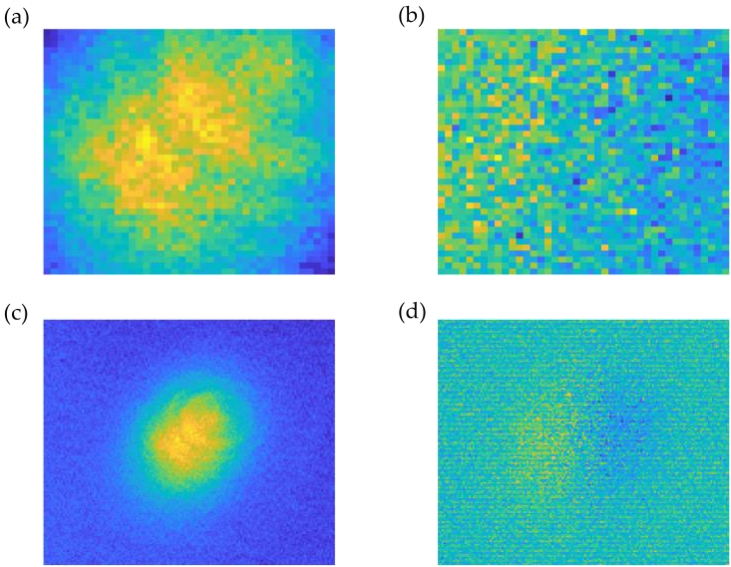

Figure 4: Figure displaying images of the first two principal components for 30 μm by 30 μm and 100 μm by 100 μm crop sizes that are discarded. Principal component 1 represents the illumination profile being removed from the set of cropped image sizes of (a) 30 µm by 30 µm and (c) 100 µm by 100 µm. Principal component 2 represents a skew in the images, which were also discarded from the set of cropped image sizes of (b) 30 µm by 30 µm and (d) 100 µm by 100 µm. All images shown have a square pixel size of $0.75 \mu m^2$.

This was followed by the reconstruction process discarding the high rank principal components 1 and 2, shown in Figure 4 for both cropped image sizes, to flatten the images while retaining the variance from the material of interest. Principal component 1 is interpreted as the illumination profile of the beam including consistent artefacts while principal component 2 is interpreted as a skew of the image towards the right, shown in Figure 4(b) and 4(d).

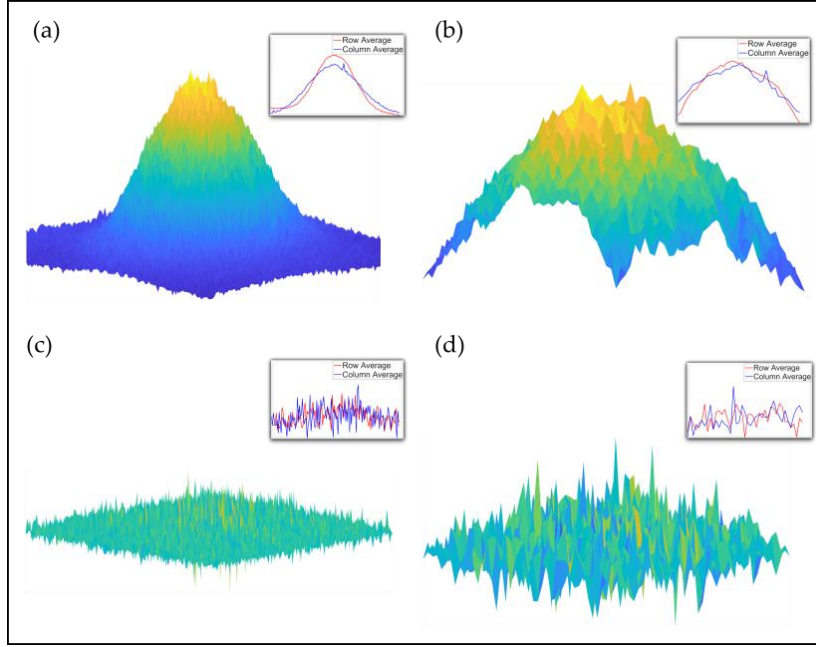

Figure 5: Surface plots of (a) an unaltered 100µm x 100µm crop, (b) an unaltered 30 µm x 30 µm crop, (c) a reconstruction of the 100 µm x 100 µm crop, and (d) a reconstruction of the 30 µm x 30 µm crop displaying the average of the scan sets when all the beams are present (Scans 1, 2, 3, 4 and 5). An inset plot of its projection onto each image axis is displayed on the upper right of each image. Scale is given as a single pixel corresponding to $0.75 \mu m^2$ for all plots shown.

The effectiveness of the technique for both crop sizes is shown in Figure 5, which shows the shape of the average of the original and the reconstructed images when all the beams are present alongside their projections onto each image axis. We note that the shape of the images is displayed in Figure 5(a) and 5(b), which is most readily attributed to the intensity profile of the probe beam, and the resulting emissions as is in principal component 1. Figure 5(a) also shows the skew of the images in the axis projections, with one of the projections tilted towards the right as replicated in principal component 2. With this information, principal components 1 and 2 are identified as the mean background profile, and the information contained in the images is more effectively reconstructed than more naïve mean background subtraction techniques. This is shown in Figure 5(c) and 5(d) with the flattened images displayed in both their surface plots and axis projections.

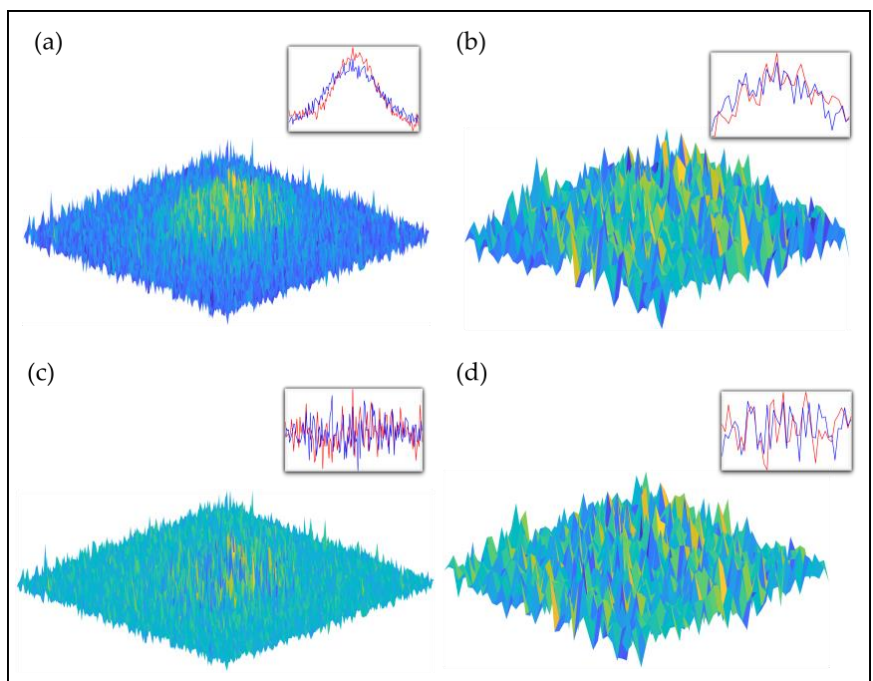

Figure 6: Surface plots with axis projections of an example CARS image displayed in the upper right for (a) an unaltered 100μm x 100μm crop of the scan frame, (b) an unaltered 30 μm x 30 μm crop of the scan frame, (c) a reconstruction of the 100 μm x 100 μm crop, and (d) a reconstruction of the 30 μm x 30 μm crop. In the reconstructed images, we make note that the images are effectively flattened in both cases, where the Gaussian illumination profile is shown in the unaltered images. The inset axis projections to the upper right of each surface plot represent the means along each row and column of the image, displayed in red and blue respectively. Scale is given as $0.75 \mu m^2$ per pixel for all images shown here.

After the reconstruction process, an example image from scan 1 and its projection onto each image axis with its reconstructed image and projection onto each image axis is shown in Figure 6. The example image is shown to be effectively flattened for both cropped image sizes of 30 μm x 30 μm and 100 μm x 100 μm respectively, indicating that the reconstruction process is effective for flattening both crop sizes individually and in average.

## 4. Conclusions

PCA has become a ubiquitous tool for background subtraction from experimental data [28, 29]. We have implemented this technique for microscopic, wide-field gaseous chemical imaging using CARS by removing the Gaussian beam profile and detector artefacts. As evidenced, this procedure successfully flattened the spatial background and isolated the signal variations. As a result, we believe this approach can improve the ability to monitor transient gas density distributions in real time and open other methods of analysis previously unavailable in the original images. In the future, this may of great relevance in industries such as cryogenics, semiconductors and processing that rely on controlled gas atmospheres where purity is essential. In this context, PCA-based background subtraction of CARS images offers a route to improve in-situ diagnostics methods [31]. This method is also consistent with previous studies demonstrating the use of PCA for noise filtering and signal enhancement in spectral imaging and microscopy [28, 32].

As much as gases have been explored theoretically, direct microscopic insights into the dynamics of IR-inactive molecules such as nitrogen and oxygen molecules have not been thoroughly investigated as compared to their IR-active counterparts. As nitrogen molecules are inactive in the infrared region, it is still challenging to study their density fluctuations dynamically in space due to extremely weak chemical signals stemming from low order Raman interactions and low concentration of gaseous samples. We believe this technique could be used to explore density micro-dynamics under various thermodynamic and reactive conditions. For example, the evolution of nitrogen or oxygen density fluctuations could be correlated during the formation of nitrates under controlled flow conditions. In microtubing, flows can have larger density gradients than in ambient air, thus providing an ideal environment for exploring the feasibility of this technique. Hence, our findings could contribute to the ability to visualize microscopic flows or other transport events of gases otherwise invisible to infrared optical techniques.

PCA has provided an extra layer of credibility to our microimaging abilities. Our findings impact optical sensing in a way that wide-field CARS chemical images can be flattened, and the molecular dynamics can be easily viewed for typical imaging procedures. With further study, it could also enable quantitative visualization of the gaseous species under wide-field illumination, revealing spatial and temporal structure in IR-inactive gas density fields. Comparable beam-correction and PCA denoising methods have been applied to precision beam profiling in interferometric measurements [33] and to combustion plume analysis [34].

**Author Contributions:** Conceptualization, G.O.A; Methodology, G.O.A; Validation, L.C.; Formal Analysis, G.O.A., L.C. and D.B.; Investigation, L.C.; Data Curation, L.C.; Writing – Original Draft Preparation, L.C., G.O.A., H.W., D.B. and M.S.; Writing – Review and Editing, D.B. and M.S.; Visualization, L.C. and D.B.; Supervision, G.O.A. and H.W.; Project Administration, G.O.A.

**Funding:** The authors declare no funding.

**Data Availability Statement:** Data will be made available if requested.

**Acknowledgments:** The authors declare no acknowledgements.

**Conflicts of Interest:** The authors declare no conflicts of interest.

**Abbreviations**

The following abbreviations are used in this manuscript:

| | |
|---|---|
| CARS | Coherent Anti-Stokes Raman Spectroscopy |
| PCA | Principal Component Analysis |
| NDIR | Non-dispersive Infrared |
| FWM | Four-Wave Mixing |
| FWHM | Full Width at Half Maxima |
| EMCCD | Electron-Multiplying Charge-Coupled Device |
| PC | Principal Component |